\documentclass[aps, pra, twocolumn, groupedaddress, superscriptaddress, floatfix, longbibliography]{revtex4-2} 

\usepackage{amsmath}
\usepackage{amssymb}
\usepackage{graphicx}
\usepackage{float}
\usepackage[english]{babel}
\usepackage{dsfont}
\usepackage{epstopdf}
\usepackage{soul}
\usepackage{color}
\usepackage{braket}
\usepackage{mathrsfs} 
\usepackage[normalem]{ulem}
\usepackage[colorlinks=true,
            linkcolor=blue,
            urlcolor=blue,
            citecolor=blue]{hyperref}

\definecolor{SMblue}{rgb}{0.1,0.1,0.9}
\definecolor{red}{rgb}{1,0,0}

 \definecolor{ASmagenta}{rgb}{1,0.2,1}

\emergencystretch=\maxdimen
\usepackage{makecell}

\begin{document}

\title{Probing the Topological Anderson Transition in Quasiperiodic Photonic Lattices via Chiral Displacement and Wavelength Tuning}

\author{Abhinav Sinha}
\affiliation{Department of Physics, Indian Institute of Science, Bangalore 560012, India}
\author{Trideb Shit}
\affiliation{Department of Physics, Indian Institute of Science, Bangalore 560012, India}
\author{Avinash Tetarwal}
\affiliation{Department of Physics, Indian Institute of Science, Bangalore 560012, India}
\author{Diptiman Sen}
\affiliation{Centre for High Energy Physics, Indian Institute of Science, Bangalore 560012, India}
\author{Sebabrata Mukherjee}
\email{mukherjee@iisc.ac.in}
\affiliation{Department of Physics, Indian Institute of Science, Bangalore 560012, India}
\date{\today}

\begin{abstract}
The interplay of topology and disorder in quantum dynamics has recently attracted significant attention across diverse platforms, including solid-state devices, ultracold atoms, and photonic systems. Here, we report on a topological Anderson transition caused by quasiperiodic 
modulation of the stronger intra-cell couplings in photonic Su-Schrieffer-Heeger lattices.
As the quasiperiodic strength is varied, the system exhibits a reentrant transition from a trivial phase to a topological phase and back to a trivial phase, accompanied by the closing and reopening of the band gap around zero energy. Unlike the traditional detection of photonic topological edge modes, we measure the mean chiral displacement from the transport of light in the bulk of the lattices. In our photonic lattices with a fixed length, the propagation dynamics is retrieved by varying the wavelength of light, which tunes the inter-waveguide couplings. 
\vspace{-1mm}
\noindent
\end{abstract}

\maketitle

\section{Introduction}\label{intro}
Topological materials exhibit remarkable robustness against defects, disorders, and imperfections~\cite{klitzing1980new, thouless1982quantized, raghu2008analogs, wang2009observation, rechtsman2013photonic, hafezi2013imaging, aidelsburger2015measuring, hasan2010colloquium, ozawa2019topological}. 
This topological robustness is usually destroyed by strong disorder, which drives a topologically non-trivial system into a trivial one. 
Interestingly, disorder can also transform a trivial system into a topological one~\cite{li2009topological, titum2015disorder, agarwala2017topological}.
The interplay of disorder and topology has opened exciting avenues, leading to {the observation of disorder-induced topological materials, known as} topological Anderson insulators (TAIs)~\cite{li2009topological, titum2015disorder}. The phenomenon was first predicted in HgTe quantum wells in the context of the quantum spin Hall 
effect, where strong disorder resulted in quantized conductance in an otherwise trivial system in the clean limit~\cite{li2009topological}. Since then, disorder-induced topology has been extensively studied, with numerical and experimental investigations exploring its underlying mechanisms~\cite{groth2009theory,Prodan_2011, Chen_2012,song2014aiii,mondragon2014topological,hsu2020topological, lin2020topological,Zhang2020,lin2021real, shi2021disorder}.

{Artificial systems, like photonic lattices and ultra-cold atoms, provide an excellent experimental platform for exploring a broad range of transport and localization} phenomena~\cite{christodoulides2003discretizing, garanovich2012light, bloch2012quantum}. These lattices enable precise engineering of coupling strengths and geometries, facilitating the realization of diverse Hamiltonians. %
Their {design flexibility,} tunability, and {the ability to directly visualize the evolution of an initial state} make them ideal for studying topological transitions and measuring invariants in complex systems. 
{In the presence of strong random disorder, topological Anderson phases have been observed in two-dimensional Floquet photonic networks~\cite{stutzer2018photonic} and in one-dimensional bipartite cold-atomic wires~\cite{meier2018observation}; see also Refs.~\cite{liu2020topological, ren2024realization, zhang2021experimental}.}

{Unlike systems with random disorder~\cite{anderson1958absence, lee1985disordered, schwartz2007transport}, quasiperiodic lattices~\cite{aubry1980analyticity, lahini2009observation, kraus2012topological, ganeshan2015nearest,roy2021reentrant,aditya2023, vaidya2023reentrant} 
with deterministic disorder exhibit intriguing localization transitions in low dimensions. The effects of different types of disorder on topological phases~\cite{song2012dependence, girschik2013topological, kuno2019disorder, tang2022topological, nakajima2021competition} and on topological phase boundaries constitute an active field of research.
Notably, a {single topological} transition from a trivial to
a non-trivial phase has been reported in one-dimensional quasiperiodic photonic circuits~\cite{longhi2020topological,gao2022observation, cheng2022observation}, by detecting the presence of zero-energy modes. 
In this work, we {measure the bulk topological invariant to} demonstrate a {\it reentrant} topological transition %
in photonic Su-Schrieffer-Heeger (SSH) lattices with quasiperiodic 
disorder in stronger intra-cell couplings.
The TAI phase in our model is gapped for all disorder realizations, i.e., the zero-energy topological edge modes spectrally reside in a {\it sizable band gap}, unlike in Refs.~\cite{longhi2020topological,gao2022observation, cheng2022observation} and \cite{ren2024realization}.
We %
realize femtosecond laser-fabricated~\cite{davis1996writing,szameit2010discrete} SSH lattices and measure the mean chiral displacement~\cite{cardano2017detection, maffei2018topological, Longhi2018, d2020bulk, roberts2022topological} 
from the output intensities %
to quantify the topological invariant.
Importantly, we employ a wavelength-tuning technique, %
which enables us to vary the normalized propagation distance by tuning the inter-waveguide couplings.
The accuracy of the wavelength tuning is first validated by observing a topological phase transition %
in `clean' SSH lattices.
By measuring the variation of the mean chiral displacement with quasiperiodic strength, we then observe the reentrant topological Anderson transition.
We also show that the topological phase diagram %
is independent of the specific irrational number used to realize quasiperiodicity. 

\begin{figure*}[hbt]
    \centering
\includegraphics[width=0.8\linewidth]{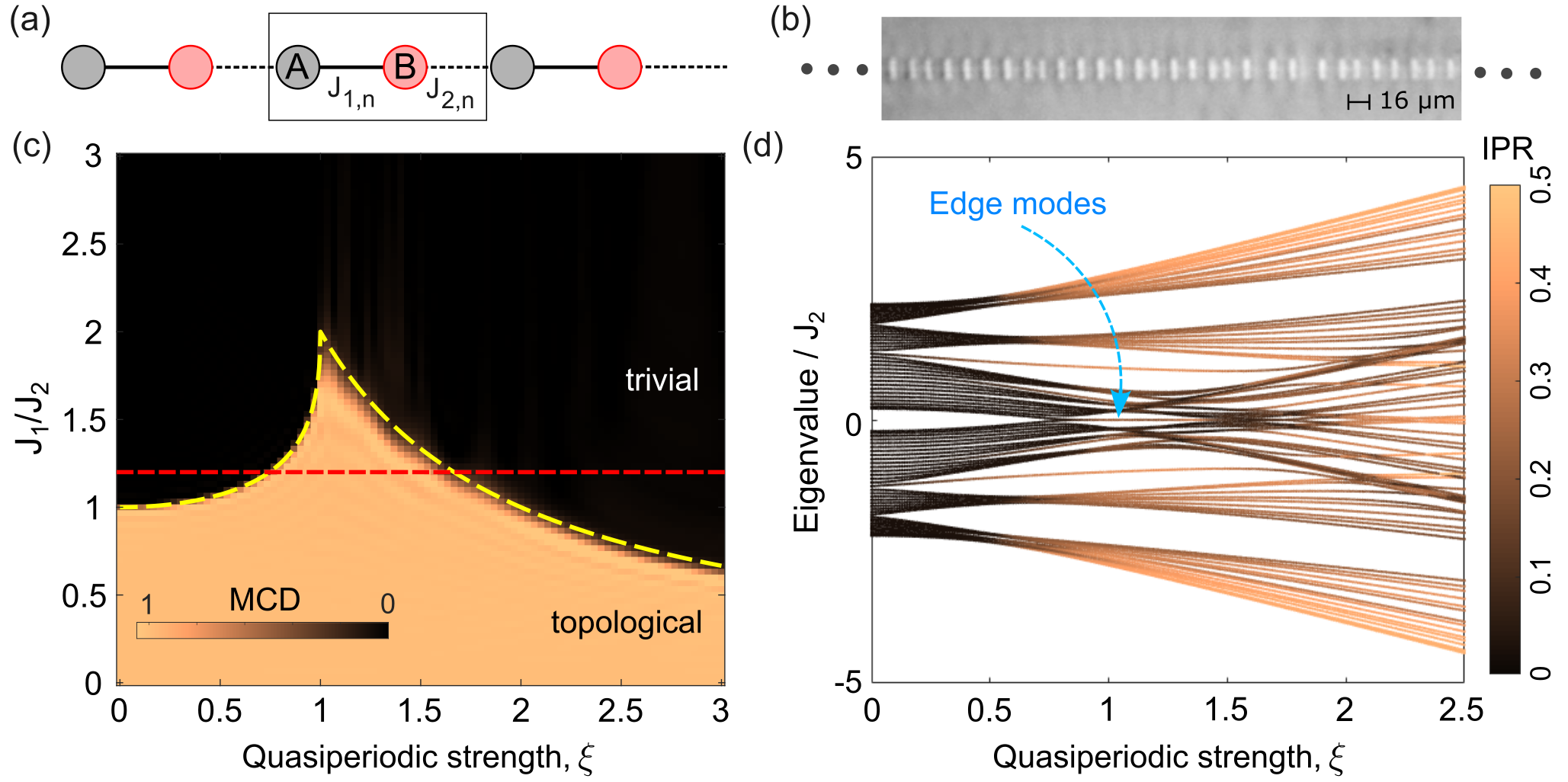}
    \caption{(a) {Schematic of a one-dimensional lattice with fixed {inter-cell} couplings $J_{2,n}$ and
    quasiperiodic {intra-cell} couplings $J_{1,n}$.
    (b) Cross-sectional} facet image of a {femtosecond} laser-written quasiperiodic lattice {($30$ out of a total of
    $80$ sites are shown here)}. 
    (c) {Mean chiral displacement} as a function of quasiperiodic strength $\xi$ and the ratio of mean {intra- to inter-cell} couplings $J_1/J_2$. {The dashed yellow line indicates the phase boundary.} (d) Energy spectrum as a function of quasiperiodic strength $\xi$ for experimentally realized system sizes and $J_1/J_2 \!=\! 1.2$ [{indicated by the dashed red line in (c)}]. Zero-energy topological edge modes within the band gap are visible {for 
    $0.75 \lesssim \xi \lesssim 1.67$.
    The} color bar indicates the inverse participation ratios (IPR) of the eigenstates. %
    }
    \label{fig_1}
\end{figure*}

The paper is organized as follows. In Section~\ref{model}, we introduce the SSH lattice model with quasiperiodic intra-cell coupling disorder. We discuss the resulting topological phase diagram using the mean chiral displacement as a measure of the topological character. In Section~\ref{Wavelength_tuning}, we present the concept of wavelength tuning.  Experimental results for the clean and quasiperiodic SSH lattices are discussed in Section~\ref{expt}. A topological phase transition in the clean SSH lattice is probed utilizing the wavelength tuning technique. We then observe transitions from a trivial to a topological phase and back to a trivial phase induced by quasiperiodic disorder. Finally, in Section~\ref{conclusions}, we emphasize the significance of this work and highlight its future implications.


\section{Model} \label{model}
In our study, we consider a {one-dimensional} quasiperiodic model described by the tight-binding Hamiltonian
\begin{eqnarray}
\hat{H} = - \sum \big( J_{1,n}\hat{a}_n^\dagger \hat{b}_n + J_{2,n} \hat{b}_{n}^\dagger \hat{a}_{n+1} + \text{H.c.} \big) \, , \label{eq1} 
\end{eqnarray} 
where $\hat{a}_n, \, \hat{b}_n$ are annihilation operators for the two sites (A and B) of the $n$-th unit cell, as shown in Fig.~\ref{fig_1}(a).
The {intra-cell} coupling $J_{1,n} \!=\! J_1(1+\xi\cos(2\pi\alpha n+\phi))$ is spatially modulated quasiperiodically with a strength $\xi$ and frequency $\alpha = (\sqrt{5}+1)/2$ (golden ratio), while the 
{inter-cell} coupling is kept fixed $J_{2,n} = J_2$.
Here, $\phi$ is the phase of the quasiperiodic pattern. 

In the case of an evanescently coupled waveguide array, the transport of optical fields is governed by the discrete Schr{\"o}dinger equation~\cite{christodoulides2003discretizing, garanovich2012light}, $i\partial_z \psi=\hat{H}\psi$, where the propagation distance $z$ plays the role of time, and $\psi$ is a column vector whose elements are the peak amplitudes of the electric field of light at the lattice sites. By adjusting the inter-waveguide spacing, the desired couplings are realized in experiments.

In the absence of quasiperiodicity (i.e., $\xi\!=\!0$), Eq.~\eqref{eq1} describes a lattice with bipartite couplings; this is known as the Su-Schrieffer-Heeger model~\cite{su1979solitons}. For $J_1<J_2$, the SSH model possesses a
non-trivial topology characterized by a non-zero 
topological invariant (called the
Zak phase~\cite{atala2013direct}) of the bulk bands, and the appearance of zero-energy edge modes for a finite system terminated with weak couplings at the two
ends.
Our model with $\xi\!\neq\!0$ can exhibit interesting topological phases, {as described below. 
Since disorder or quasiperiodicity breaks the translational symmetry of the system, the traditional topological invariants defined in the
reciprocal $k$-space are not useful in our case.
Additionally, many real-space variants~\cite{bianco2011mapping, mondragon2014topological, lin2021real} of such topological invariants depend on bulk eigenstates, making}
direct experimental observation challenging. %
To address this, we use the mean chiral displacement (MCD) {\cite{cardano2017detection, maffei2018topological, Longhi2018, d2020bulk, roberts2022topological}, which converges to the winding number in the limit of a 
long propagation length.} %
{The chiral displacement is defined as
\begin{eqnarray}
C(z) = \langle2\Gamma X\rangle = 2 ~\sum_n ~ n~ (|\psi^{A}_{n}|^2-|\psi^{B}_{n}|^2)\, , \label{chiral-displacement}
\end{eqnarray}  
where} $\Gamma$ is the chiral symmetry operator, $X$ is the unit cell operator, 
and $\psi^{A}_{n}$ ($\psi^{B}_{n}$) is the wave function at the A (B) site of the $n$-th unit cell.
We {initialize the input state at a single}
site {of the $0$-th unit cell at the middle of the lattice} and record $C$ as it evolves. Averaging the chiral displacement over a long propagation length $L$ converges it to the \textit{local winding number} (local with respect to the input unit cell)
\begin{eqnarray}
\bar{C} = \lim_{L\rightarrow\infty}\frac{1}{L}\int_0^{L} {\text{d}}z ~C(z). \label{mean-chiral-displacement}
\end{eqnarray}
In the case of %
disordered systems, further averaging over multiple input sites within the bulk, leading to $\langle\bar{C}\rangle$, reveals the %
{topological invariant}
of the system in the limit of large system size \cite{bianco2011mapping,meier2018observation}  (here, $\langle\rangle$ denotes the unit cell average). %

Considering a large system size of $200$ unit cells and %
a long propagation distance, we numerically obtain the mean chiral displacement 
as a function of $\xi$ and $J_1/J_2$, as shown in Fig.~\ref{fig_1}(c). %
In the absence of any disorder, the $J_1\!<\!J_2$ ($J_1\!>\!J_2$) region in Fig.~\ref{fig_1}(c) corresponds 
to the topologically non-trivial (trivial) phase of the SSH model. %
Upon introducing a quasiperiodic disorder, it is expected that, up to a certain disorder strength, %
the model should remain topologically non-trivial %
for $J_1\!<\!J_2$. 
An interesting reentrant transition occurs in the $J_1\!\gtrsim\!J_2$ region, where phase 
transitions from %
trivial to non-trivial and back to trivial {occur %
as the quasiperiodic strength is increased}. 
A specific case of $J_1/J_2 = 1.2$, indicated by the dashed red line in Fig.~\ref{fig_1}(c), is realized experimentally (see later). Considering the experimentally realized parameters, we plot the eigenvalues of the system with respect to the quasiperiodic strength $\xi$, as shown in Fig.~\ref{fig_1}(d). 
The spectrum is symmetric around the zero eigenvalue due to the chiral symmetry. In the clean case, we observe a band gap, but as %
$\xi$ increases, the gap closes and reopens, signaling a topological phase transition. Specifically, for %
$0.75 \lesssim \xi \lesssim 1.67$, zero-energy topological modes emerge. 
Importantly, introducing quasiperiodic disorder in the stronger intra-cell couplings $J_{1, n}$ (instead of the weaker inter-cell couplings $J_{2, n}$) makes our model, spectrum, and the nature of the topological phase diagram distinct from Ref.~\cite{longhi2020topological,gao2022observation, cheng2022observation}, where the zero-energy modes in the TAI phase are {\it gapless}. Indeed, in 
the TAI phase of our system, the zero-energy edge modes lie in a bulk gap which is about $0.3\,J_2$ for
all disorder realizations. 
This also contrasts with the TAI phase found in Ref.~\cite{ren2024realization}, where the bulk gap is about $10^{-6} ~J_2$ or less and depends significantly on the disorder realization.

The energy of the topological edge modes in Fig.~\ref{fig_1}(d) remains fixed at zero, and they consistently appear as edge modes, even when we vary the frequency $\alpha$ (over irrational values) and phase $\phi$ of the quasiperiodic pattern.
We analytically study the topological phase diagram by probing the zero-energy edge modes (Appendix~\ref{app:topophasediagram}) and confirm that the phase boundary is, in fact, independent of the choice of $\alpha$ (as long as it is irrational) and $\phi$ in the limit that the system is semi-infinite (i.e., the chain has one end). {Indeed, by identifying the condition under which zero-energy edge modes can exist, we obtain the following} analytical formula for the phase boundary, {indicated by the dashed yellow line in Fig.~\ref{fig_1}(c),} %
\begin{eqnarray}
\frac{J_1}{J_2} ~=~ 
\begin{cases}
    ~\dfrac{2}{1+\sqrt{1-\xi^2}}& ~~\text{if } ~\xi\leq 1, \\[0.5cm]
    ~2/\xi              & ~~\text{if }~ \xi>1.
\end{cases}\label{eq:boundary_formula_maintext}
\end{eqnarray}
This result is elegant in the sense that $J_2/J_1$ (the inverse of Eq. \eqref{eq:boundary_formula_maintext}) 
is linear with $\xi$ for $\xi>1$ and forms a quarter ellipse for $\xi\leq1$. 
The colorbar in Fig.~\ref{fig_1}(d) denotes the inverse participation ratio (IPR) {of the eigenstates}, which is 
defined as 
\begin{eqnarray}
\label{IPR}
\text{IPR} = \frac{\sum_{n}|\psi^{A}_{n}|^4+|\psi^{B}_{n}|^4}{(\sum_{n} |\psi^{A}_{n}|^2+|\psi^{B}_{n}|^2)^2} \, ,
\end{eqnarray}
where $n$ ranges over the unit cell indices.
{The IPR is a measure of localization -- it reaches unity when the wave function is localized to a single site.} %
{Note that the system primarily hosts delocalized eigenstates for small values of $\xi$; see Fig.~\ref{fig_1}(d).} {As a function of the quasiperiodic strength, we observe three phases of delocalized,
mixed and localized eigenstates; for details, see more Appendix~\ref{app:IPR}.} 
{We also note that the spatial profile of the zero-energy edge modes in the TAI phase varies with $\phi$. Since a single site initial state on the edge A site may not efficiently excite the edge modes, the measurement of $\langle\bar{C}\rangle$ is a natural experimental choice.}

\begin{figure*}[hbt]
    \centering
\includegraphics[width=0.9\textwidth]{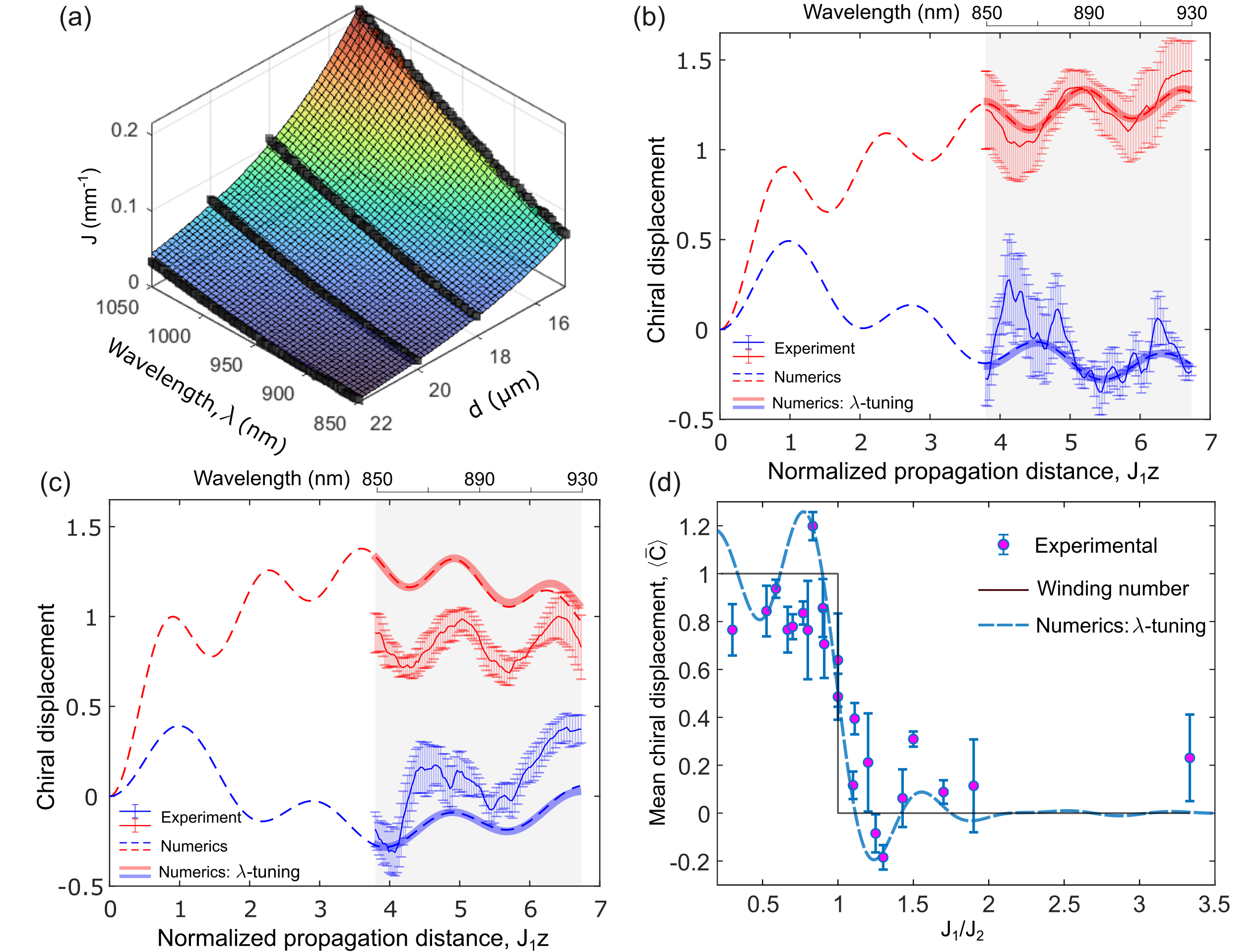}
\caption{(a) Wavelength tuning:~coupling strength $J$ as a function of wavelength and inter-waveguide spacing $d$. Black squares denote the experimental data and the surface is its curvefit. 
(b) The chiral displacement 
$\langle C \rangle$,
averaged over four independent measurements, for two different values of  $J_1/J_2 = 0.83$ (red) and $1.30$ (blue). The dashed and thick solid lines are obtained numerically from Eq.~\eqref{chiral-displacement} and wavelength tuning, respectively.
(c) Same as (b) for $J_1/J_2 = 0.77$ (red) and $1.42$ (blue). The background shade marks the wavelength tuning range of $80$~nm.
(d) Mean chiral displacement, $\langle \bar{C}\rangle$, as a function of $J_1/J_2$ for the SSH lattices, showing a topological phase transition at $J_1/J_2\!=\!1$.
Numerically calculated MCD considering wavelength tuning and experimentally realized system parameters is shown by the blue dashed line. The error bars in (b-d) are standard error of the mean value. }
\label{fig_wt}
\end{figure*}

\section{Wavelength Tuning} \label{Wavelength_tuning}
The measurement of the mean chiral displacement requires probing intensity profiles as a function of $z$.
In our experiments, the maximal propagation length of the photonic lattice is fixed, and measurements are often limited to the output intensities, although %
top imaging techniques have been employed in certain studies~\cite{szameit2007quasi} to capture propagation dynamics in one-dimensional lattices. 
Here, we consider a %
wavelength-tuning technique that enables the extraction of the dynamics of light intensity by tuning the couplings with the wavelength $\lambda$ of the incident light.
The idea is straightforward to explain for a one-dimensional lattice with a homogeneous coupling $J$. In this case, when light is coupled at the $0$-th site far away from the edges, the state at the $n$-th site is given by~\cite{jones1965coupling} $\psi_n(z)\!=\!(i)^n \mathcal{J}_n(2Jz)$, where $\mathcal{J}_n$ is the $n$th-order Bessel function of the first kind. Evidently, the dynamics is determined by $Jz$, and hence, one can tune the wavelength of light to vary $J$, instead of varying $z$, to probe the dynamics. 

The wavelength-tuning technique can be extended to the quasiperiodic lattice as long as the coupling ratios $J_{1, n}/J_2$ remain wavelength-independent. 
We can express the Hamiltonian in Eq.~\eqref{eq1} as $\hat{H}\!=\!J_2 \hat{H'}$, where $\hat{H'}$ is independent of $\lambda$ if $J_{1, n}/J_2$ does not change within the wavelength range of interest. In that case, the evolution operator is given by $\hat{U}(z)=\exp(-i \hat{H'} J_2 z)$. Evidently, the propagation dynamics of the system is described by a normalized propagation distance $J_2z$.
This approach leverages the fact that the output intensities remain identical for any combination of $J_2$ and $z$ that yields the same value of $J_2 z$. 
%


\section{Experiments} \label{expt}
The photonic devices are fabricated using the femtosecond laser writing technique~\cite{davis1996writing, szameit2010discrete} in borosilicate (Corning Eagle XG) glass; see also Appendix~\ref{app:fabrication}. The inter-waveguide evanescent coupling is estimated by characterizing light transport in a set of two-waveguide devices.
Experimental (black data points) and fitted variation of coupling with wavelength and inter-waveguide spacing $d$ is shown in Fig.~\ref{fig_wt}(a). The coupling varies almost linearly with $\lambda$ and exponentially with $d$~\cite{szameit2007control, mukherjee2017dissipatively}. By tuning the laser wavelength from $850$~nm to $1050$~nm, the normalized propagation distances can be varied from $3.76$ 
to $12.7$ in our experiments for a $76.2$~mm-long device with a fixed inter-waveguide spacing of $16\, \mu$m. However, to minimize the deviation in the coupling ratios, we use a wavelength range of 
$80$~nm in experiments, as described later. Utilizing the variation in couplings with waveguide spacing and wavelength $J(d, \lambda)$, all propagation calculations are performed by solving the discrete Schr{\"o}dinger equation.
We note that the wavelength-tuning method also works for two-dimensional lattices where top imaging is more challenging. %

\begin{figure*}[hbt]
    \centering
    \includegraphics[width=0.92\linewidth]{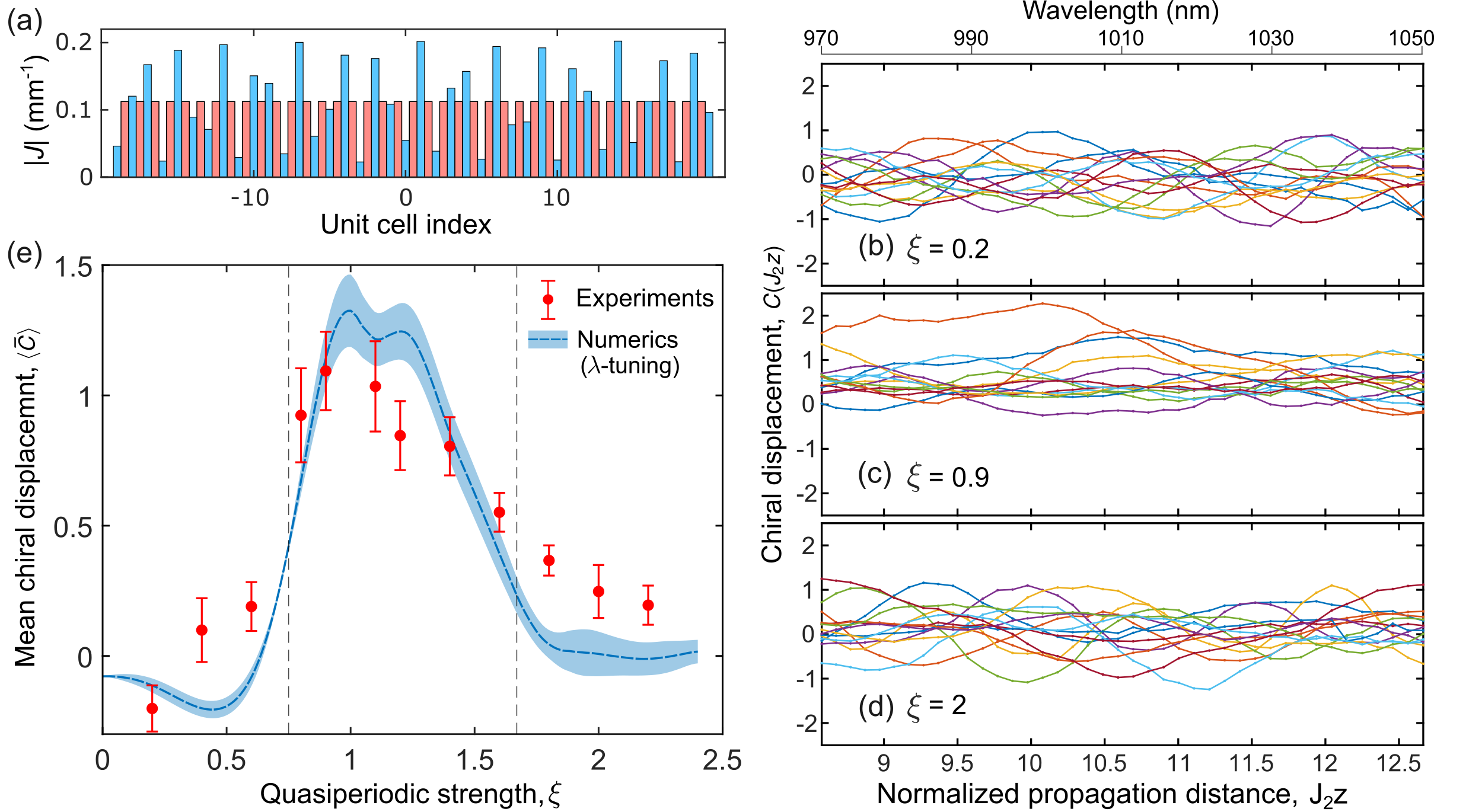}
    \caption{(a) Experimentally realized quasiperiodic couplings. Here, blue (red) color indicates $J_{1, n}$ $(J_{2, n})$ at $970$~nm wavelength of light. (b-d) Experimentally obtained chiral displacements $C(J_2 z)$ as a function of wavelength (or, $J_2z$)  for quasiperiodic strengths $\xi = 0.2$, $\xi = 0.9$ and $\xi = 2$, respectively. Different coloured lines represent {initial excitation at} different input sites. 
    (e) Mean chiral displacement as a function of $\xi$. Red circles are experimental data points with error bars denoting the standard error of the mean value of $\langle \bar{C}\rangle$. 
    The blue dashed line is obtained numerically for experimentally realized parameters, and the shade indicates the standard error for $51$ realizations of the phase $\phi$ of the quasiperiodic pattern. The vertical dashed lines mark the topological phase transition points in the thermodynamic limit.
    }
    \label{fig_3}
\end{figure*}

To demonstrate the effectiveness of the wavelength-tuning method, we designed and fabricated $11$ sets of SSH lattices, each consisting of $40$ waveguides. 
%
For these lattices, we set $\xi\!=\!0$, $J_1\!=\! 0.088$~mm$^{-1}$ at $\lambda\!=\!930$~nm, and varied $J_2$ from $0.028$ to $0.161$~mm$^{-1}$.
Horizontally polarized light was launched at a single {A site} {of a unit cell in the bulk}, and the output intensity patterns were measured within the wavelength range of $850$~nm to $930$~nm. The upper limit of $\lambda\!=\!930$~nm is chosen so that the output state does not reach the edge of the lattice. 
The chiral displacement, averaged over 
four independent measurements 
$\langle C \rangle$,
is plotted in Fig.~\ref{fig_wt}(b) for two different values of 
$J_1/J_2 = 0.83$ (red) and $1.30$ (blue). 
%
%
The dashed lines, obtained from $z$-evolution (i.e., by solving Eq.~\eqref{chiral-displacement}), agree well with the wavelength tuning results shown by the thick solid lines in Figs.~\ref{fig_wt}(b).
Another data set is shown in Fig.~\ref{fig_wt}(c) for $J_1/J_2\!=\!0.77$ (red) and $1.42$ (blue).
Evidently, the chiral displacement shows distinct values in the topologial and trivial phases.
The fluctuations in these measurements arise from unavoidable small random disorder $(\Delta J/J \approx \pm 3\%)$ in the lattices. 
%
%
In these bulk dynamics measurements, we can denote A (B) sites as B (A) sites and redefine the couplings, to obtain the chiral displacement for both $J_1/J_2$ and its inverse from the same lattice. 
We then obtain the mean chiral displacement by averaging the measured chiral displacement over the considered wavelength range (i.e., normalized propagation distance).
%
The MCD as a function of $J_1/J_2$, clearly showing a topological phase transition at $J_1/J_2=1$, is presented in Fig.~\ref{fig_wt}(d).
Numerically calculated MCD considering wavelength tuning and experimentally realized system parameters is shown by the bluish dashed line in Fig.~\ref{fig_wt}(d). The dark line in the same figure indicates the the winding number, i.e., MCD in the thermodynamic limit. 

{To investigate bulk transport and topological phase transitions in our quasiperiodic model, we fabricated $12$ sets of photonic lattices varying the quasiperiodic strengths from $\xi\!=\!0.2$ to $2.4$. 
It is worth noting that the} {intra-cell} coupling $J_{1,n}$ can become negative for higher values of $\xi$; however, we can gauge out the negative couplings and effectively work with $|J_{1,n}|$, which is feasible for fabrication. %
Indeed, in a one-dimensional tight-binding model, altering the phases of the coupling constants does not impact the physical observables of the system; {see 
Appendix~\ref{app:GaugeTransformation}. 
Figure~\ref{fig_3}(a) presents the quasiperiodic couplings realized in the experiment with $\xi\!=\!0.8$}
{at $970$~nm wavelength.}
We calculate the chiral displacement $C(J_2z)$ from the extracted dynamics to assess the topological properties. For instance, Figs. \ref{fig_3}(b-d) present $C(J_2z)$ as a function of the wavelength (or $J_2z$) for three values of {$\xi\!=\!0.2, \, 0.9$, and $2$, respectively; the different} colors 
are associated with fourteen measurements by coupling light at fourteen different input sites. 
In these experiments,  
the variation of the coupling ratio distribution $|J_{1, n}|/J_2$ is not significant
for a wavelength span of 
of interest; see Appendix~\ref{wavelength_tuning}.
By averaging over the input sites and the normalized propagation distance, we obtain the mean chiral displacement $\langle \bar{C} \rangle$, as shown in Fig.~\ref{fig_3}(e). 
The blue dashed line shows the numerically obtained MCD for the experimentally realized parameters, averaged over $51$ realizations of the phase $\phi \in [-\pi, \pi]$ of the quasiperiodic pattern. The shaded region indicates the standard error. The experimental data set agrees well with the numerical results.
%

The agreement between experimental and theoretical results validates our approach and demonstrate the feasibility of probing topological properties through wavelength tuning in photonic lattices.
Indeed, a reentrant topological Anderson transition is clearly visible in Fig.~\ref{fig_3}(e). 
However, due to finite system size in our experiments, the MCD values are not quantized and sharp phase transitions are not observed. As discussed in Appendix~\ref{wavelength_tuning}, the precision of our results can be further enhanced by employing a longer sample with more lattice sites. 
%
Additionally, we note that the presence of random disorder in 
the on-site energy and a next-nearest-neighbor coupling can destroy the chiral symmetry. However, these effects are insignificant in our experiments and do not significantly affect the measurement of the MCD. %
%


\section{Conclusions} \label{conclusions}
We have experimentally and numerically studied light transport in the bulk of a quasi-periodic photonic SSH lattice. By recording the output intensity patterns as a function of the wavelength of %
light, we determine the mean chiral displacement that captures the topological features of the system. Importantly, we demonstrate a reentrant transition from a trivial phase to a gapped topological phase and back to a trivial phase as a function of the quasiperiodic strength.
Our results provide key insights into the physics of topology and quasiperiodic disorder and showcase the versatility of photonic lattices as a platform for topological studies.
Additionally, it should be noted that laser-fabricated photonic lattices are a natural platform to realize Floquet topological materials and study the influence of self-focusing Kerr nonlinearity using intense laser pulses~\cite{mukherjee2020observation, eisenberg1998discrete, shit2025intensity}. Evidently, our results will be useful for further exploring the interplay of {periodic driving,} topology, disorder and nonlinear interactions.

\begin{acknowledgments}
We thank Ferdinand Evers and Tapan Mishra for useful discussions. S.M.~gratefully acknowledges support from the Indian Institute of Science (IISc) for funding through a start-up grant, 
the Ministry of Education, Government of India, for funding through STARS (MoE-STARS/STARS-2/2023-0716) and the Infosys Foundation, Bangalore. 
A.S. thanks the Department of Science \& Technology (DST) for INSPIRE-KVPY fellowship.
T.S.~and A.T.~thank IISc and CSIR, respectively, for their Ph.D. scholarships. 
D.S.~thanks SERB, India, for funding through Project No.~JBR/2020/000043. We sincerely thank Nicholas Smith of Corning Inc.~for providing high-quality glass wafers.
\end{acknowledgments}

\section*{DATA AVAILABILITY}
The data supporting this study’s findings are available within the article. Raw files can be made available upon reasonable request.

\appendix

        \counterwithout{equation}{section}
        \renewcommand{\theequation}{A\arabic{equation}}%
        \setcounter{equation}{0}
         \setcounter{section}{0}
         \renewcommand{\thesection}{\Alph{section}}%



\section{Topological phase diagram}\label{app:topophasediagram}
{Figure~\ref{fig_1}(c) in the main text shows the mean chiral displacement as a function of $J_1/J_2$ and $\xi$.}
Here, we 
derive an analytical { expression for the phase boundary. Our goal is to identify the condition under which zero-energy edge modes can exist, as the topological phase with the correct termination can support such modes.}
{For the quasiperiodic SSH lattice, we have the} following 
equations in the site basis for an eigenmode with energy $E$,
\begin{eqnarray}
    E\psi^{A}_{n} &=& J_2\psi^{B}_{n-1} ~+~ J_{1,n}\psi^{B}_{n} \, ,\label{TB1}\\
    E\psi^{B}_{n} &=& J_{1,n}\psi^{A}_{n} ~+~ J_2\psi^{A}_{n+1} \label{TB2} \, .
\end{eqnarray}
{Unlike the main text, we will consider here a semi-infinite lattice, where the unit cell index $n$ varies from 1 to infinity.}
{For the zero-energy eigenstate, Eqs.~(\ref{TB1}, \ref{TB2}) imply
that %
}
\begin{eqnarray}
    \psi^{A}_{n} = (-1)^{n-1}\frac{\prod_{i=1}^{n-1} J_{1,i}}{J_2^{n-1}}\psi^{A}_{1} \, ,
\end{eqnarray}
\begin{eqnarray}    
    \psi^{B}_{n} = 0\, .
\end{eqnarray}
This shows that for $E=0$, the amplitudes on all the {} $B$-sites vanish, while the {$A$-site amplitudes} follow a recursion relation leading them to be proportional to the { amplitude on edge} A site, $\psi_1^A$. %
For 
a state localized near the edge, the amplitude must decay as $n\rightarrow\infty$, leading to the condition
\begin{eqnarray}\label{eq:Q}
\frac{\prod_{n=1}^{N} J_{1,n}}{J_2^{N}}<1\;\quad \text{as\quad} N\rightarrow\infty \, ,
\end{eqnarray}
{where the total number of sites is $2N$.} Substituting 
{the couplings $J_{1,n}$ and $J_{2,n}$ in Eq.~\eqref{eq:Q}, we obtain the phase diagram presented in Fig.~\ref{fig_Q_phase_diagram},
which is in} excellent agreement with %
Fig.~\ref{fig_1}(c). This reinforces the fact that the topological phase reported using mean chiral displacement corresponds to zero-energy edge modes.
\begin{figure}[htbp]
    \centering
    \includegraphics[width=0.85\linewidth]{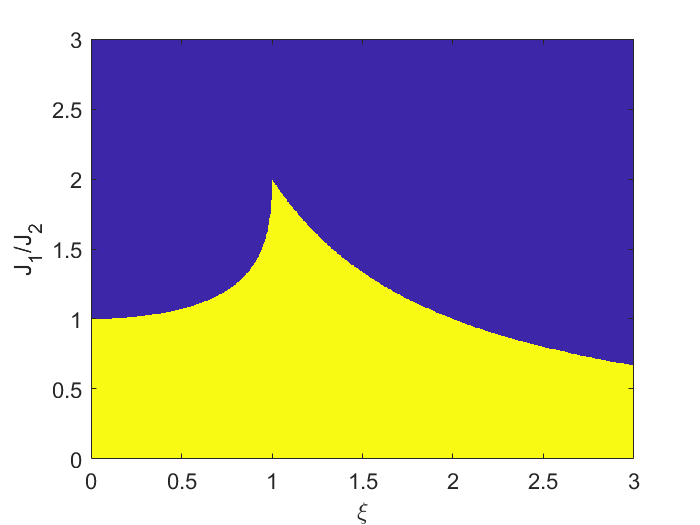}
    \caption{Topological phase diagram for a large system size of 2000 unit cells. The yellow region satisfies the condition \eqref{eq:Q}. 
    }
    \label{fig_Q_phase_diagram}
\end{figure}

{For the quasiperiodic pattern considered here,
we now show that the shape of the phase boundary is independent of the choice of the irrational number $\alpha$ and the phase $\phi$ for a very large system size.} %
The phase boundary 
satisfies the following critical condition
\begin{eqnarray}
\frac{\prod_{n=1}^{N} J_{1,n}}{J_2^{N}}&=&1 \, ,
\end{eqnarray}
which implies
\begin{eqnarray}
\frac{1}{N}\sum_{n=1}^{N}\log |1+\xi\cos(2\pi\alpha n+\phi)| &\!=\!& \log\left(\frac{J_2}{J_1}\right). \label{eq:sum_logJ}
\end{eqnarray}
Note that the term $(2\pi\alpha n+\phi)$ [mod$ ~2\pi$] %
uniformly samples all the points in $[0,2\pi]$ 
as $N\! \rightarrow \! \infty$, because of the {\it irrationality} of $\alpha$. This allows us to convert the sum in the LHS of Eq.~\eqref{eq:sum_logJ} to an integral,
\begin{eqnarray}
\log\left(\frac{J_2}{J_1}\right) = \int_{0}^{1} {\text{d}}x ~\log |1+\xi\cos(2\pi x)|.
\end{eqnarray}
The integral can be evaluated using complex variables and yields the
expression
\begin{eqnarray}
\frac{J_1}{J_2} = 
\begin{cases}
    ~\dfrac{2}{1+\sqrt{1-\xi^2}} & \text{if } ~\xi\leq 1, \\[0.5cm]
    ~2/\xi            & \text{if } ~\xi>1.
\end{cases}\label{eq:boundary_formula}
\end{eqnarray}
Eq.~\eqref{eq:boundary_formula} accurately describes the phase boundary and, notably, is independent of the choice of phase $\phi$ and $\alpha$ as long as it is irrational. The dashed yellow curve in Fig.~\ref{fig_1}(c) corresponds to the phase boundary defined by this equation.

\section{Gauging out negative couplings} \label{app:GaugeTransformation}
In this section, we will explain the transformation which allows us to take the absolute value of the couplings to eliminate negative values. Consider a general one-dimensional tight-binding model
\begin{eqnarray}
    H = -\sum_n ~[t_n \hat{c}^\dagger_n \hat{c}_{n+1}+ \text{H.c.}],
\end{eqnarray}
where $t_n$ is the coupling constant, and $\hat{c}_n$ represents the annihilation operator at site $n$.
Now, suppose that we change the phase of any particular coupling
\begin{eqnarray}
    t_j \rightarrow t_j ~\exp(i\theta)\;.
\end{eqnarray}
We can then apply either of the following transformations
\begin{eqnarray}
    c_{j'}\rightarrow c_{j'} ~\exp(i\theta) ~\text{ for all }~ j'\leq j,
\end{eqnarray}
\begin{eqnarray}
\text{or } ~~c_{j'}\rightarrow c_{j'} ~\exp(-i\theta) ~\text{ for all } ~ {j'} \geq j+1,
\end{eqnarray}
to remove the phase $\theta$ from $t_j$, while keeping the Hamiltonian unchanged. Thus, these transformations allow negative values of couplings to be gauged out, enabling us to work with the absolute values of the couplings in experiments. %

\section{Localization properties of the\\ quasiperiodic model}\label{app:IPR}

In this section, we analyze the localization properties of the eigenstates in our quasiperiodic model for a fixed coupling ratio given by $J_1/J_2=1.2$. In addition to the observed topological transitions, the system undergoes a localization transition, which we characterize using the inverse participation ratio (IPR). Our results reveal three distinct phases: delocalized, mixed, and fully localized states (Fig. \ref{fig_A_IPR}).

For quasiperiodic strength $\xi\lesssim0.5$, the system primarily hosts delocalized eigenstates. Beyond this, in the range 
$0.5\lesssim\xi\lesssim1.75$, the eigenstates exhibit a coexistence of localized and delocalized characteristics, forming a mixed phase. When $\xi\gtrsim1.75$, the system fully transitions into a localized phase, where all eigenstates are confined with IPR $\gtrsim 0.1$.

\begin{figure}[htbp]
    \centering
    \includegraphics[width=1\linewidth]{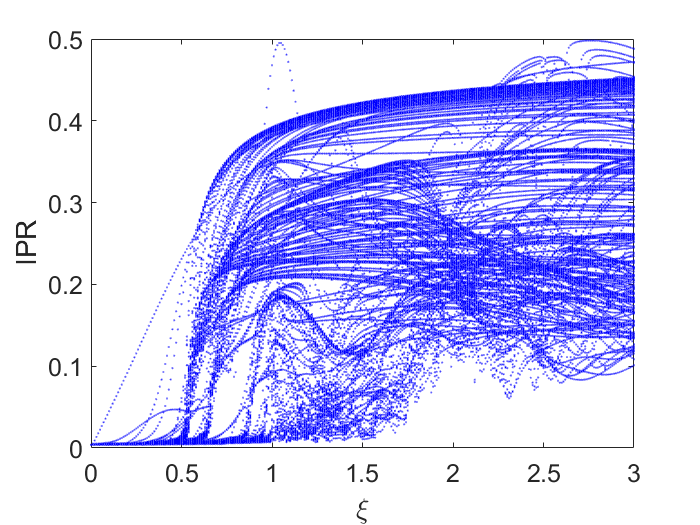}
    \caption{IPRs of the eigenstates of the quasiperiodic model for $J_1/J_2=1.2.$
    }
    \label{fig_A_IPR}
\end{figure}

\begin{figure}[t]
    \centering
    \includegraphics[width=0.8\linewidth]{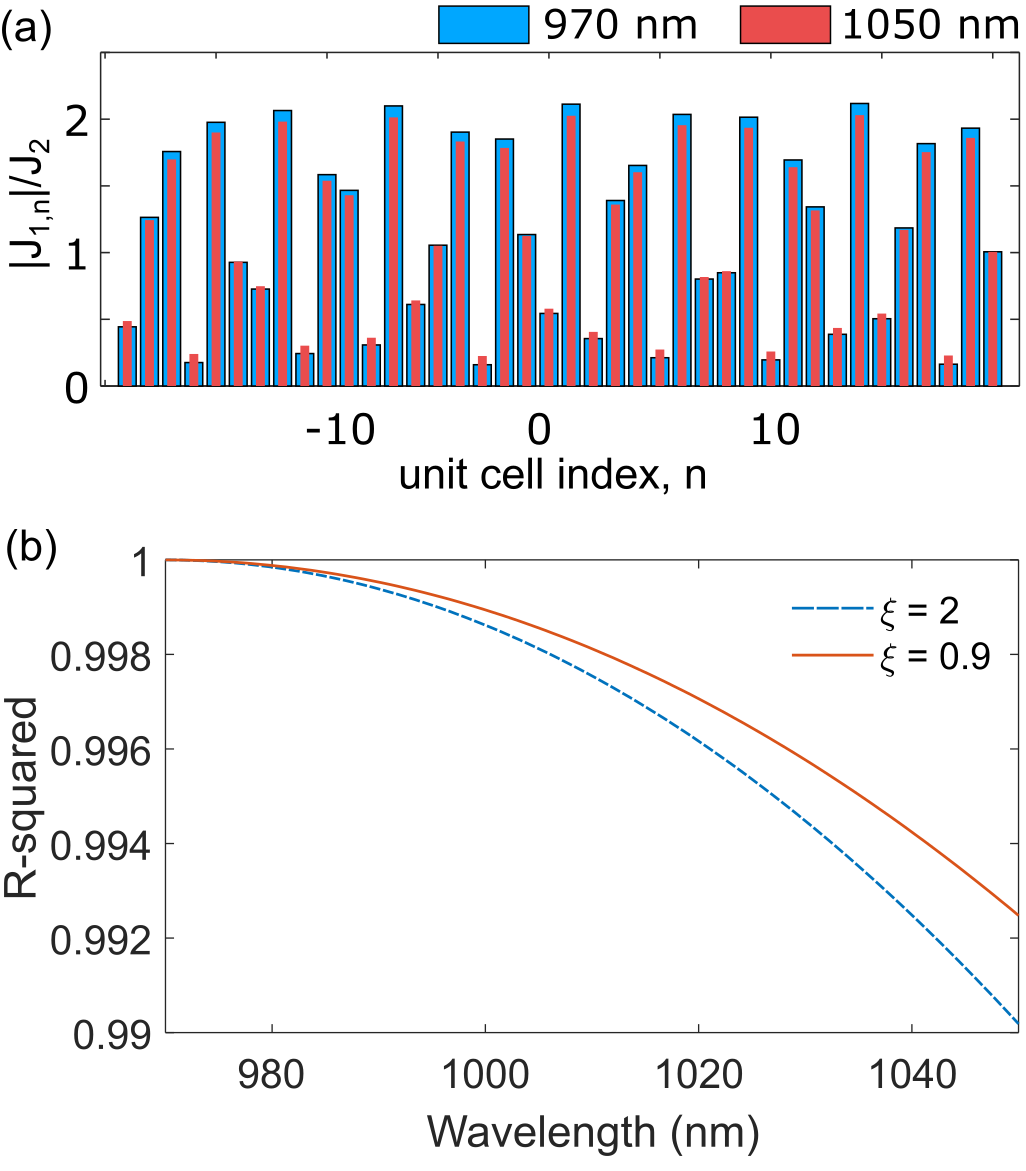}
    \caption{Experimentally measured coupling ratio $|J_{1, n}|/J_2$ for two different wavelengths $\lambda\!=\!970$~nm and $1050$~nm at $\xi\!=\!0.9$. The overall variation of the coupling ratio distribution is not significant.
    (b) R-squared value obtained from the measured $|J_{1, n}|/J_2$ as a function of the wavelength of light, with $970$~nm as the reference, indicating that its variation is small in this wavelength range. The blue line shows the R-squared value for a larger disorder strength of $\xi=2$.
    }
    \label{fig_A_wavelengthTuning}
\end{figure}

\section{Effectiveness of wavelength tuning}\label{wavelength_tuning}

In this appendix, we examine both clean and quasiperiodic SSH lattices, demonstrating the accuracy with which the MCD can be extracted using the wavelength tuning technique, especially for large system sizes.

Ideally, the wavelength-tuning technique is effective when the coupling ratio $J_{1, n}/J_2$ remains constant over the wavelength range considered. 
To assess this, we experimentally measure the inter-waveguide coupling $J(\lambda, d)$ as a function of wavelength $\lambda$ and inter-waveguide spacing $d$. 
Fig.~\ref{fig_A_wavelengthTuning}(a) presents the coupling ratio at two different wavelengths for our quasiperiodic lattice with $\xi\!=\!0.9$. Within the experimental wavelength range ($970$~nm to $1050$~nm), the variation in the coupling ratio distribution is minimal, as further confirmed by the R-squared values shown in Fig.~\ref{fig_A_wavelengthTuning}(b)
For this calculation, we used $|J_1|/J_2$ at $970$~nm as a reference and computed the R-squared value of the coupling ratio distribution as a function of $\lambda$. In Fig.~\ref{fig_A_wavelengthTuning}(b), we also plot the R-squared value for  a larger disorder strength $\xi\!=\!2$.
Evidently, across the entire disorder range considered in our experiments, the coupling ratio $|J_1|/J_2$ does not vary significantly. 

\begin{figure}[]
    \centering
    \includegraphics[width=0.9\linewidth]{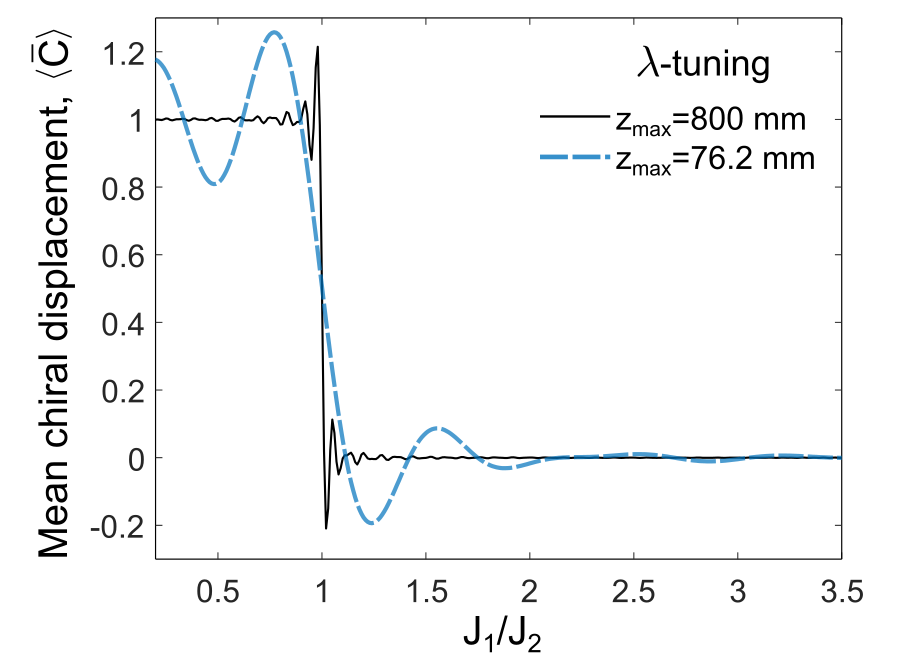}
    \caption{The blue dashed line is the same as shown in Fig.~\ref{fig_wt}(d), calculated for the finite system size realized in experiments. The dark line is the MCD for a larger system size ($N=400$ and $z_{\text{max}}\!=\!800$~mm), showing sharp topological phase transitions.}
    \label{fig_7_ssh_clean}
\end{figure}

Figures~\ref{fig_wt}(d) and \ref{fig_3}(e) in the main text
show numerically calculated and experimentally obtained MCD for our finite lattice systems realized in the experiments. By considering a larger system size and longer propagation distances, we can achieve nearly quantized MCD values with a sharp change at the topological phase transition points. 
To illustrate this, we first examine a clean SSH lattice consisting of $N \!=\! 400$ unit cells with a propagation length of $z_{\text{max}}\!=\!800$~mm. 
Using the same wavelength range as in our experiments (i.e., $850$–$930$~nm), we observe highly quantized MCD values across the $J_1/J_2$ range, with a sharper transition from $1$ to $0$, as shown by the dark line in Fig.~\ref{fig_7_ssh_clean}.
It should be highlighted that the state recycling technique described in Ref.~\cite{mukherjee2018state} can be useful to experimentally realize such a long propagation distance.

\begin{figure*}[]
    \centering
    \includegraphics[width=0.85\linewidth]{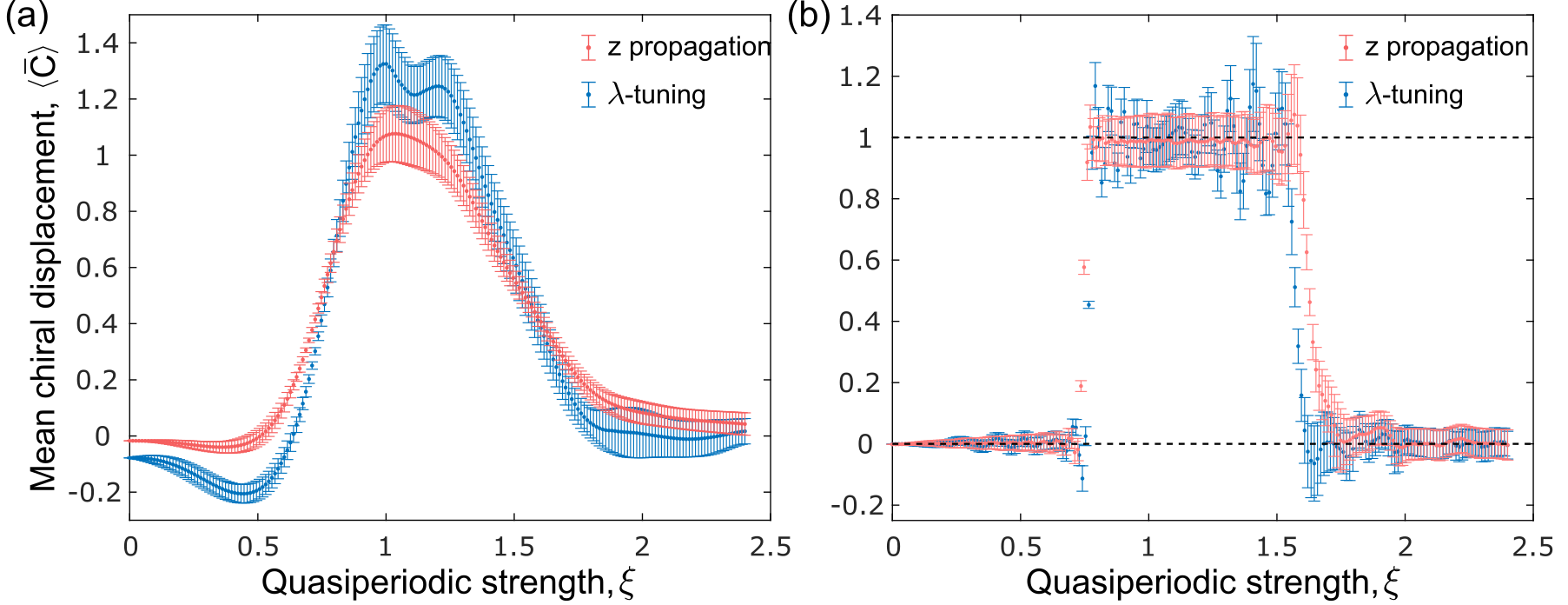}
    \caption{(a) Mean chiral displacements obtained from $z$-evaluation, (i.e., Eq.~\eqref{mean-chiral-displacement}) and wavelength tuning for $z_{\text{max}}\!=\!76.2$~mm-long quasiperiodic SSH lattices. (b) Same as (a) for $N\!=\!400$ unit cells and $z_{\text{max}}\!=\!800$~mm propagation distance. Notice that the sharpness of the topological phase transition becomes evident for longer propagation. For each value of $\xi$, the MCD is obtained averaging over $51$ quasiperiodic disorder realizations. The error bar indicates standard error in MCD.}
    \label{fig_8_ssh_quasi}
\end{figure*}

To investigate topological Anderson transitions, we consider quasiperiodic lattices with experimentally realized  as well as larger syatem sizes (i.e., $400$ unit cells and $800$~mm propagation length).
We calculate the MCD using both $z$-evolution and wavelength tuning. 
For each quasiperiodic strength, the MCD value was numerically obtained by averaging over $51$ realizations of the phase $\phi \in [-\pi, \pi]$ of the quasiperiodic pattern. 
We present the numerical results for $z_{\text{max}}\!=\!76.2$~mm and $800$~mm-long devices in Figs.~\ref{fig_8_ssh_quasi}(a, b), respectively. A reentrant topological Anderson transition is clearly visible in Fig.~\ref{fig_8_ssh_quasi}(a), however, the phase transitions are not sharp. Additionally, the MCD values in the topological phase deviates from the integer value of $0$ and $1$. This is expected as MCD coverges to the winding number in the limit of long propagation, see Eq.~\ref{mean-chiral-displacement}. In case of a longer propagation distance of $800$~mm in Fig.~\ref{fig_8_ssh_quasi}(b), the wavelength tuning data shows good agreement with the $z$-evolution data. Importantly, notice the sharp topological transitions with nearly quantized MCD values. We note that the error bars in Fig.~\ref{fig_8_ssh_quasi} indicate standard deviation in MCD obtained for $51$ disorder realizations. 
Evidently, the longer propagation significantly enhances the quantization of MCD values and sharpens the phase transitions compared to the shorter propagation case.

The MCD obtained from wavelength tuning can slightly differ from the value calculated from the $z$-evolution. For example, the second topological transition point at $\xi\!=1.67$ is observed to deviate to $\xi\!=1.62$ in Fig.~\ref{fig_8_ssh_quasi}(b). This small effect arise because the coupling ratio $|J_{1, n}|/J_2$ does not hvae a ``strictly fixed value". In future works, these issues can be addressed by designing waveguide lattices with insignificant variation in coupling ratio  in the wavelength range of interest.

\section{Fabrication and characterization details}\label{app:fabrication}
In this Appendix, we briefly discuss how the photonic lattices were fabricated and characterized. All waveguide-based photonic devices were created using femtosecond laser writing \cite{davis1996writing,szameit2010discrete} -- a laser-based technique that induces refractive index modifications within a transparent dielectric medium. %
We use a Yb-doped fiber laser system (Satsuma, Amplitude) to generate $260$ fs optical pulse trains with 500 kHz repetition rate and $1030$ nm central wavelength. 
To regulate the power and polarization of the laser beam, we utilized a polarizing beam splitter and wave plates. 
The laser beam with circular polarization is focused (using a $0.4$ NA lens) within a Corning Eagle XG glass wafer mounted on high-precision $x$-$y$-$z$ translation stages (Aerotech Inc).
Each waveguide is created by translating the glass wafer once through the focus of the laser beam at a speed of $6$~mm/s and an average laser power of $170$~mW.

The photonic devices were characterized using a wavelength tunable super-continuum source (NKT Photonics). We focused the light at a single desired waveguide at the input and imaged the output intensity pattern on a CMOS camera. For broadband performance, we used achromatic doublet lenses. We used horizontally polarized light for all characterization experiments.


We estimate the propagation loss of the waveguides by subtracting the measured coupling loss from the insertion loss. 
As the wavelength of light is tuned from $850$~nm to $1050$~nm, the propagation loss increases from $0.19$ to $0.40$~dB/cm. 
In our experiments, all the lattice sites have the same propagation loss for a given wavelength of light. In other words, we do not have any site dependent loss.
In this situation, we can renormalize the intensities as $|\psi_j(z)|^2/\sum_j|\psi_j(z)|^2$ to obtain the dynamics of optical intensity. Consequently, the mean chiral displacement remains unaffected by the experimentally measured propagation losses.



%

\end{document}